\documentclass[10pt,journal,compsoc,onecolumn]{IEEEtran}

\ifCLASSOPTIONcompsoc
  \usepackage[nocompress]{cite}
\else
  \usepackage{cite}
\fi

\usepackage{amssymb}
\usepackage{graphicx}
\usepackage{algorithm}
\usepackage{algorithmic}
\usepackage{amsmath}
\usepackage{amsthm}
\usepackage{booktabs}
\usepackage{multirow}

\ifCLASSINFOpdf
\else
\fi
\begin{document}

\title{A Multi-Sensor Fusion Parking Barrier System with Lightweight Vision on Edge}

\author{Yuwen Zhu, Feiyang Qi, Zhengzhe Xiang
\IEEEcompsocitemizethanks{
\IEEEcompsocthanksitem JY. Zhu, F. Qi, and Z. Xiang are with the School of Computer Science and Computing, Hangzhou City University, Hangzhou, China. 
}
}

\markboth{Journal of \LaTeX\ Class Files,~Vol.~14, No.~8, August~2015}%
{Shell \MakeLowercase{\textit{et al.}}: Bare Advanced Demo of IEEEtran.cls for IEEE Computer Society Journals}

\IEEEtitleabstractindextext{%
\begin{abstract}
To address the challenge that detection accuracy, edge-side real-time performance, low-power operation, and end-to-end business linkage are difficult to satisfy simultaneously in parking scenarios, this paper proposes and implements an intelligent parking barrier system based on deep learning and multi-sensor fusion. The system adopts a three-layer collaborative architecture consisting of an edge sensing node layer, a cloud business service layer, and a front-end management application layer. On the edge side, a Raspberry Pi 5 serves as the core controller and integrates a camera, infrared ranging sensor, MPU6050 attitude sensor, and LoRa module to complete parking-space state sensing and local decision-making. On the cloud side, a Spring Boot backend provides unified multi-space management, order billing, alarm handling, and heartbeat monitoring. On the application side, a Vue3+Vite front end implements parking-map rendering, state visualization, and hierarchical alarm display. At the algorithmic level, because the parking task only concerns a single target class (\texttt{Vehicle}), YOLOv3-tiny is structurally pruned, the detection-head output channel is set to 18, and model weights are compressed to approximately 33 MB. At the decision level, an asymmetric infrared-vision-inertial fusion state machine is designed. Through an “infrared trigger — visual confirmation — inertial fallback” mechanism, robustness is improved under nighttime, occlusion, and impact-disturbance conditions. Experimental results show that after more than 5000 training iterations, the loss decreases from approximately 10.0 and stabilizes at about 0.6, while late-stage mAP@0.5 reaches 96.5\%--98.2\%. On Raspberry Pi 5, single-frame inference latency is approximately 600--850 ms at 416x416 input resolution, meeting the polling requirements of 5 s when idle and 10 s when occupied. Average system power consumption decreases from 4.02 W to about 1.02 W, yielding approximately 74\% energy savings. Joint debugging tests further verify the comprehensive advantages of this solution in detection accuracy, response timeliness, energy control, and engineering deployability.
\end{abstract}

}

\maketitle

\IEEEdisplaynontitleabstractindextext

\IEEEpeerreviewmaketitle

\ifCLASSOPTIONcompsoc
\IEEEraisesectionheading{\section{Introduction}\label{sec:introduction}}
\else
\section{Introduction}
\label{sec:introduction}
\fi

\subsection{Research Background}
In recent years, rapid economic development and substantial improvements in living standards have led to a continuous increase in the number of motor vehicles. Statistical reports indicate that although parking infrastructure has expanded steadily, the newly added spaces still fail to close the widening demand gap. The contradiction between existing parking supply and actual parking demand remains prominent in China, especially in large cities and commercial centers \cite{ref1}.

\begin{figure}[H]
\centering
\includegraphics[width=0.8\textwidth]{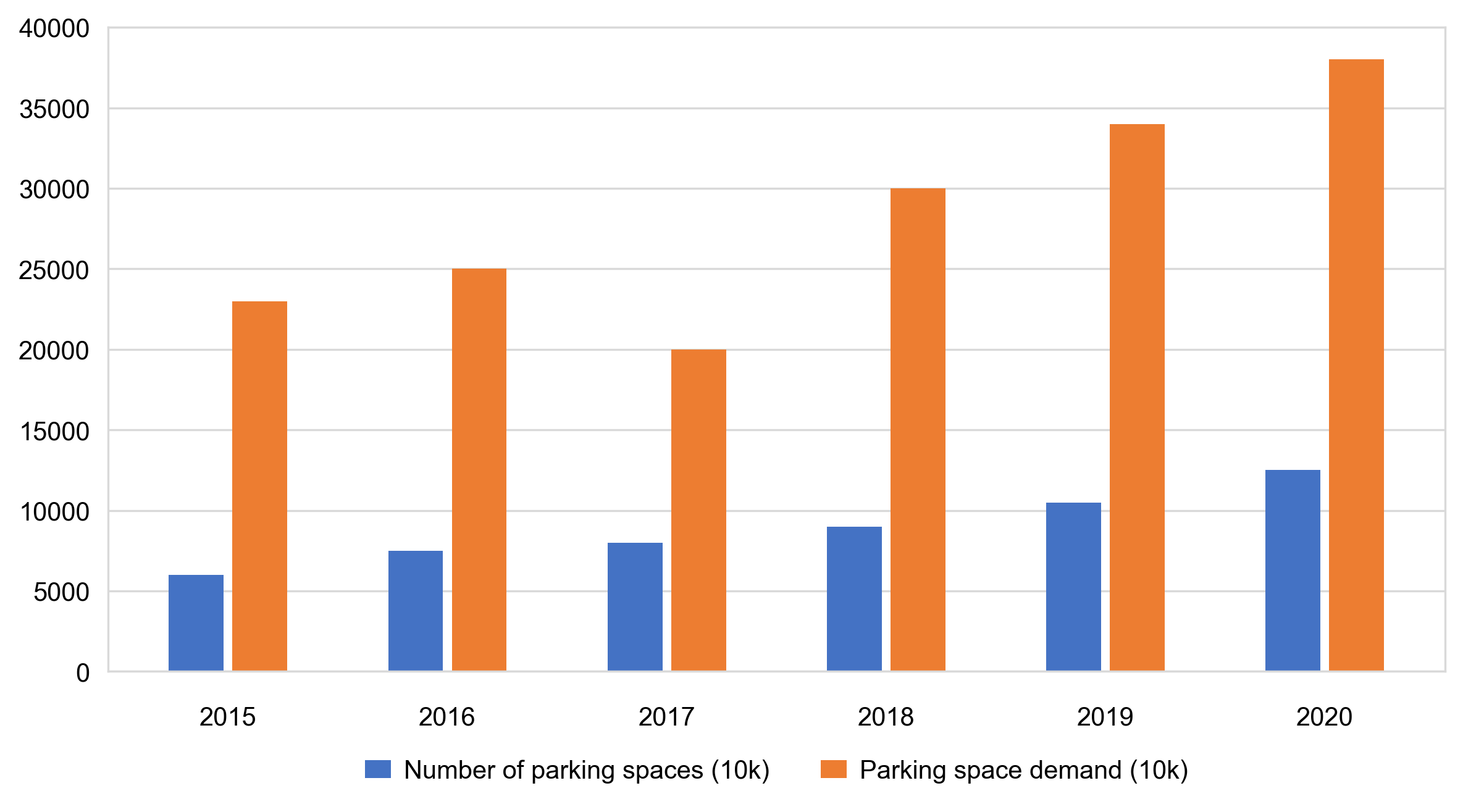}
\caption{Comparison Between Existing Parking Supply and Parking Demand in China}
\label{fig:parking_demand}
\end{figure}

As shown in Fig. \ref{fig:parking_demand}, the gap between parking demand and existing built parking capacity has expanded significantly in recent years. This supply-demand imbalance not only leads to severe “parking difficulty” in urban core areas, but also highlights the urgency of improving space turnover and refined operations through intelligent technologies. Under high parking pressure, conventional parking-lot management methods can no longer satisfy efficient operation requirements. At present, occupancy perception in many parking spaces still relies on manual inspection, geomagnetic sensing, or single photoelectric sensors. Manual recording is inefficient and costly, and is not conducive to centralized statistics and integrated management. Traditional hardware sensing solutions also suffer from limited information dimensions, high false-positive rates, and insufficiently fine-grained state interpretation, making them inadequate for real-time space-level supervision and refined billing \cite{ref2}. Recent comprehensive surveys on smart parking systems highlight the growing convergence of IoT, computer vision, and artificial intelligence as key enablers for next-generation parking management \cite{ref16}.

To overcome these technical bottlenecks, demand for intelligent parking management systems has become increasingly urgent. In recent years, deep-learning-based object detection has achieved substantial progress in tasks such as vehicle recognition and parking occupancy analysis. In particular, lightweight detectors such as the YOLO family have shown a favorable speed-accuracy trade-off \cite{ref3,ref4}. However, in practical parking-barrier deployment, pure vision solutions still face three key challenges. First, edge devices such as Raspberry Pi 5 have constrained compute and memory, making direct deployment of large and complex vision models difficult. Recent studies on FPGA-based acceleration have demonstrated significant potential for compressing YOLOv3-tiny models while maintaining detection accuracy \cite{ref13,ref21}. Second, complex environmental disturbances—including weak nighttime illumination, rain, fog, occlusion, pedestrians, and non-motor vehicles—can substantially increase the risk of false detection in single-modality vision systems. Third, there is often a lack of high-reliability, low-overhead, and anti-jitter communication links and consistency mechanisms between front-end perception and backend business systems, weakening the overall business linkage. Emerging edge-IoT frameworks have begun addressing these challenges through integrated multimodal pipelines that preserve data locality and ensure real-time operation \cite{ref20}.

In summary, a single technical approach cannot fully solve fine-grained parking-space management in complex environments. Therefore, constructing a systematic solution that combines high-reliability multi-modal recognition, low-power edge operation, and edge-cloud collaborative management is of significant engineering and practical value for intelligent parking management.

\subsection{Research Questions and Objectives}
This study focuses on “high-reliability parking occupancy determination and business linkage on low-compute platforms” and addresses several key questions: how to achieve practical visual detection speed on a Raspberry Pi 5 CPU platform, how to reduce misjudgment caused by nighttime conditions, backlighting, and non-vehicle disturbances, and how to efficiently integrate edge-side results into backend business and visualization workflows. Comparative studies of infrared and ultrasonic sensors have shown that each modality has distinct limitations, motivating the need for fusion-based approaches \cite{ref22}. Based on these questions, the objective is to establish a reusable, deployable, and maintainable edge-cloud integrated technical path that reaches an engineering-acceptable balance among accuracy, latency, energy consumption, and stability.

\subsection{Main Contributions}
The main contributions of this paper are as follows. A three-layer edge-cloud collaborative system architecture for parking barriers is designed to realize closed-loop management of parking-space state, alarms, and billing. A YOLOv3-tiny single-class pruning scheme is proposed and combined with a dedicated training strategy to achieve lightweight yet high-accuracy recognition. An infrared non-monotonic interval correction method and a visual-inertial redundant decision mechanism are introduced to improve robustness in complex scenarios. A hierarchical power-state transition strategy is established and validated via joint debugging for energy efficiency and system stability. Finally, full-process experimental results are provided, covering model training, edge deployment, and business-level joint debugging, offering practical engineering references for similar parking IoT systems.

\section{Related Work}
\subsection{Progress in Parking Vehicle Detection}
Existing parking-vision research mainly follows two directions. One direction emphasizes detection accuracy by adopting deeper networks, feature pyramids, or multitask frameworks to improve representation in complex scenes \cite{ref1,ref3,ref7}. The other direction emphasizes edge deployment capability by reducing inference latency and resource usage through lightweight backbones, model pruning, and hardware-aware acceleration \cite{ref6,ref8}. Recent advancements have demonstrated the effectiveness of model compression techniques, with structured pruning achieving up to 98.2\% model size reduction for YOLOv3-tiny while maintaining detection performance \cite{ref13,ref21}. Additionally, systematic evaluations of YOLO variants on resource-constrained platforms have provided valuable insights into the trade-offs between accuracy and inference speed \cite{ref20}. In general, the former performs well in offline evaluation but has higher deployment barriers on low-power platforms such as Raspberry Pi 5; the latter is more engineering-oriented but often exhibits reduced robustness under complex disturbances.

\subsection{Limitations of Existing Methods}
Parking-barrier scenarios usually focus only on limited states such as occupancy or non-occupancy, and generic multi-class models introduce unnecessary classification overhead, which constrains edge-side real-time performance \cite{ref5,ref6}. Meanwhile, pure vision approaches are unstable under extreme lighting and occlusion, while pure infrared approaches are vulnerable to false judgments in close-range non-monotonic regions; therefore, single-modality methods are difficult to operate stably over the long term \cite{ref2,ref11}. Comparative analyses of infrared and ultrasonic sensors reveal that environmental factors such as temperature and airflow can significantly affect measurement accuracy, necessitating multi-modal fusion strategies \cite{ref14,ref22}. In addition, relatively few studies provide end-to-end validation that simultaneously covers perception algorithms, communication protocols, backend business logic, and visualized alarms; reproducible system-level solutions remain insufficient \cite{ref4,ref12}. Building on these observations, this paper presents a unified design from three aspects: model lightweighting, multi-sensor fusion, and edge-cloud linkage.

\section{Overall System Architecture and Method}
\subsection{Three-Layer System Architecture}
The system consists of three layers with a unified data model for state transition. The edge sensing node layer uses a Raspberry Pi 5 integrated with an HD camera, infrared sensor, MPU6050, and LoRa module for local sensing and preliminary decision-making. The cloud business service layer, built with Spring Boot, is responsible for unified multi-space management, billing, device heartbeat monitoring, and alarm flow control. The front-end management application layer, implemented with Vue3 and Vite, provides parking maps, color-coded states, and highlighted alarms. In this architecture, the edge layer prioritizes real-time state inference, the cloud layer prioritizes business consistency, and the front end provides interpretable visualization and operations interaction. This division reduces single-point complexity and improves scalability and maintainability. The edge-cloud collaborative paradigm aligns with recent work on end-multi-edge-cloud computation offloading, which has demonstrated significant improvements in task completion rates and energy efficiency.

\subsection{Edge-Side Visual Inference Pipeline}
Given that the selected edge device (Raspberry Pi 5 with a 2.4 GHz Cortex-A76 processor) has no discrete GPU and limited compute capacity, the system uses the \texttt{OpenCV DNN} interface \texttt{cv2.dnn.readNetFromDarknet} to load YOLOv3-tiny and executes forward inference optimized for CPU and NEON instructions.

\begin{figure}[H]
\centering
\includegraphics[width=0.8\textwidth]{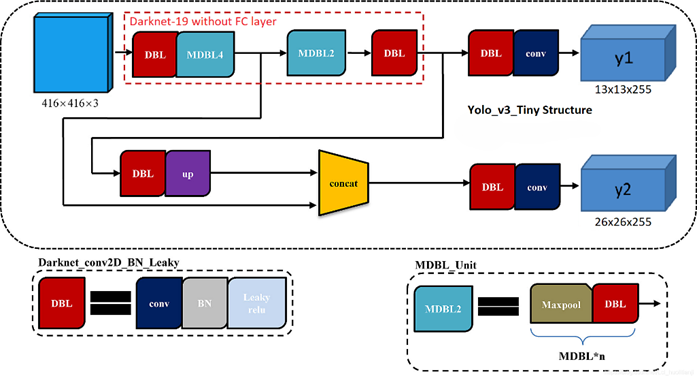}
\caption{YOLOv3-Tiny Network Architecture}
\label{fig:yolo_structure}
\end{figure}

As shown in Fig. \ref{fig:yolo_structure}, the YOLOv3-Tiny architecture adopted in this system is significantly simplified relative to the standard version. The model primarily stacks deep convolution and pooling layers, removes redundant feature-pyramid branches, and preserves backbone feature extraction capability for medium and large targets. This lightweight design significantly reduces model parameters and computational cost, effectively balancing representation capability and execution constraints, and is therefore suitable for real-time inference on low-power edge hardware such as Raspberry Pi 5.

Based on this architecture, the edge-side inference pipeline includes image acquisition, scale normalization, DNN forward pass, confidence filtering, NMS suppression, and target-state output. This study sets \texttt{CONF\_THRESHOLD=0.25}, \texttt{NMS\_THRESHOLD=0.50}, and an input resolution of 416x416 to balance recall and false positives. Under relatively fixed parking-camera viewpoints, this configuration maintains stable recognition performance and satisfies periodic edge polling requirements. Recent evaluations of YOLO models with pixel-wise region-of-interest selection have achieved over 99\% accuracy in parking occupancy detection, demonstrating the effectiveness of such approaches \cite{ref17}.

\subsection{Multi-Sensor Fusion and State Machine}
To address single-sensor misjudgment, an asymmetric fusion state machine is designed based on the principle of “low-cost sensing for trigger, high-cost vision for confirmation, and inertial information for anomaly fallback.” Visual recognition is awakened when the infrared measured distance is less than 80 cm. YOLO then outputs vehicle confidence and updates the occupancy state. If infrared is triggered but visual confidence is insufficient and MPU6050 detects an impact, a conservative “collision-induced parking” decision is made. When the tilt angle exceeds 25°, an abnormal tilt event is directly reported.

\begin{figure}[H]
\centering
\includegraphics[width=0.8\textwidth]{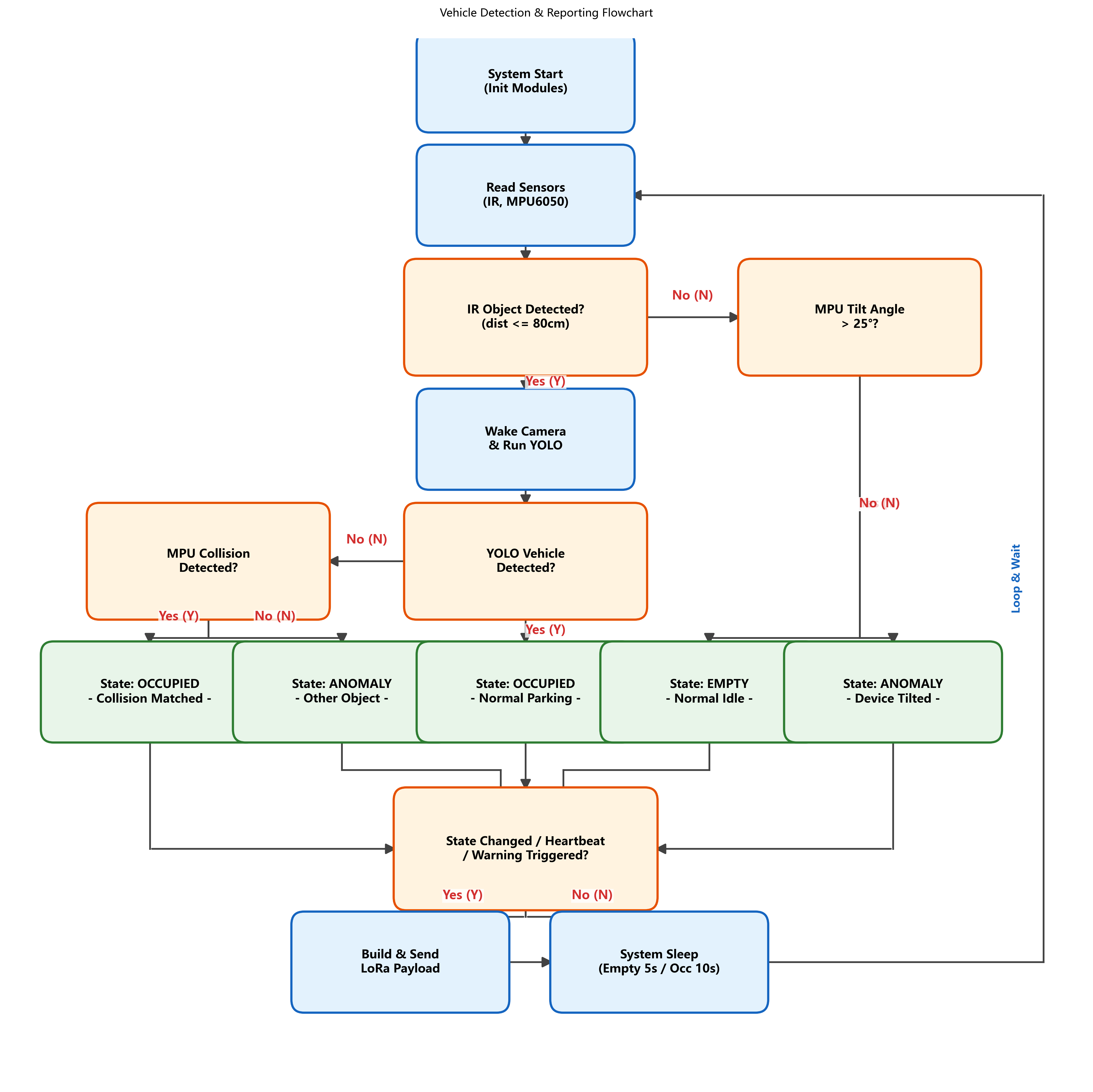}
\caption{Flowchart of the Multi-Sensor Fusion State-Machine Algorithm}
\label{fig:fusion_flowchart}
\end{figure}

In addition, to address non-monotonic infrared mapping near close-range dead zones, this work introduces bidirectional interpolation lookup and historical reading smoothing to constrain and correct transient outliers. As illustrated by the flow in Fig. \ref{fig:fusion_flowchart}, this mechanism reduces false negatives and false positives in weak-light, occlusion, and occasional impact scenarios, thereby improving the continuity and reliability of state determination. The fusion of multiple sensor modalities aligns with established multi-sensor information fusion approaches for parking space detection, which combine laser ranging, inertial measurements, and visual data to enhance robustness \cite{ref19}. Quantitative accuracy evaluations of infrared sensors for parking systems have shown that proper calibration and multi-sensor integration can significantly reduce false detection rates \cite{ref14}.

\subsection{Edge-Cloud Data Path and State Consistency Mechanism}
Edge nodes send JSON messages via LoRa serial communication. Core fields include parking-space ID, terminal ID, occupancy status, confidence score, occupancy reason, infrared distance, attitude angle, and estimated power consumption. The cloud backend adopts an idempotent state-machine update strategy for repeated reports and prioritizes newer records with higher confidence. To prevent transient jitter from triggering incorrect business events, alarm thresholds and minimum persistence windows are configured to ensure convergence among “perception state,” “business state,” and “display state.”

\section{Model Training and Parameter Settings}
\subsection{Dataset and Annotation}
Training uses the public \texttt{car\_dataset\_1class} dataset, which covers multiple vehicle forms such as sedans and SUVs.

\begin{figure}[H]
\centering
\includegraphics[width=0.8\textwidth]{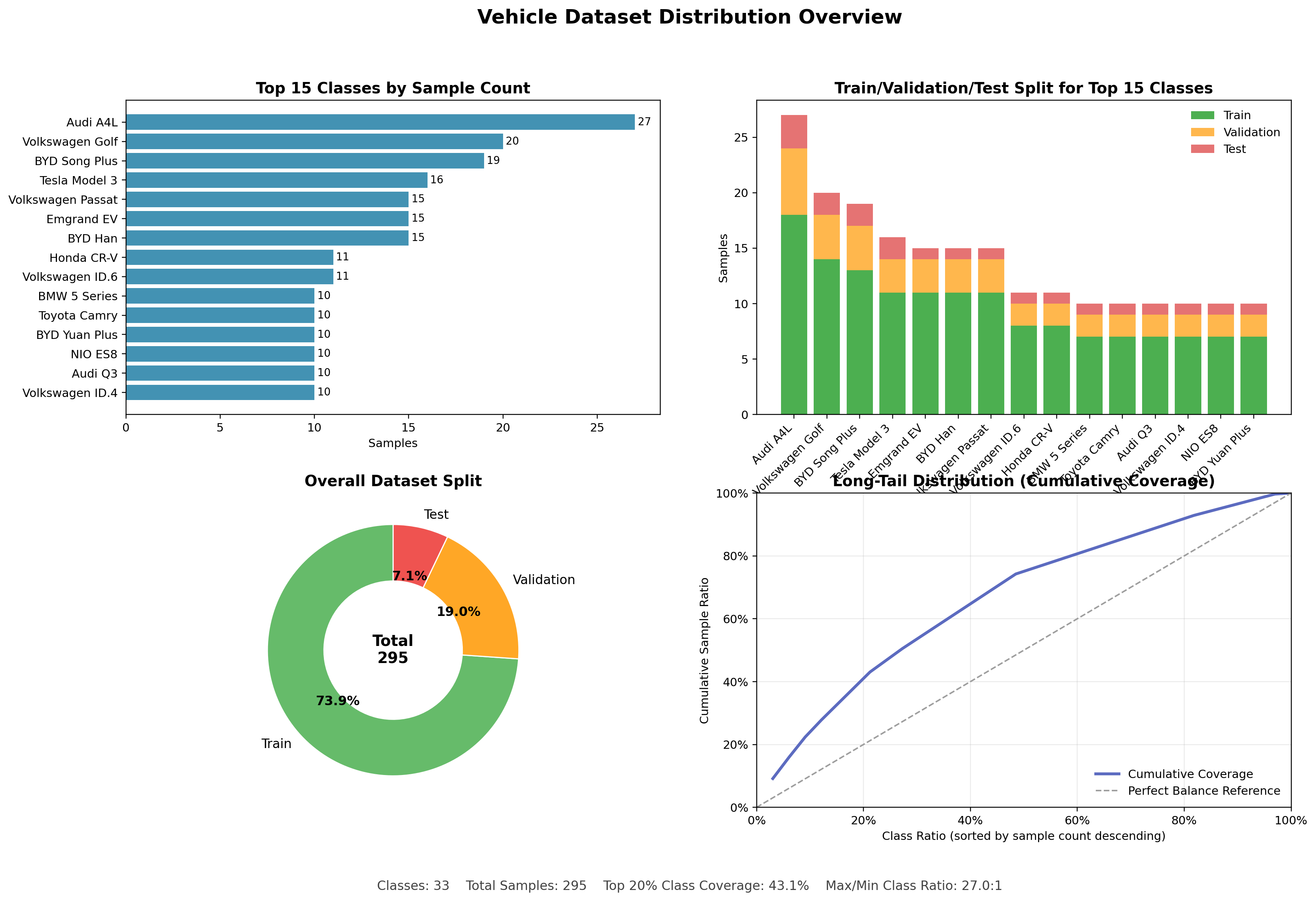}
\caption{Overall Distribution of the Vehicle Recognition Dataset}
\label{fig:dataset_dist}
\end{figure}

As shown in Fig. \ref{fig:dataset_dist}, the dataset comprehensively covers common vehicle forms (e.g., sedans, SUVs, and light passenger vehicles) and maintains substantial diversity and balance across different lighting conditions, viewing angles, and background environments. Such rich sample distribution helps the model learn more generalizable and robust visual features. Since the core business objective of parking barriers is merely to determine whether parking has occurred, all vehicle subtypes are uniformly annotated as a single class, \texttt{Vehicle}, which aligns with business requirements and avoids training ambiguity introduced by multi-class overlap. Data augmentation includes random cropping, hue/saturation jitter, and brightness perturbation to simulate shadows, backlighting, and weather changes, improving model adaptability to non-ideal parking-lot imaging conditions \cite{ref9,ref10}.

\subsection{Single-Class Model Pruning}
The original YOLOv3-tiny is designed for 80 COCO classes. This study reconfigures it for a single class according to parking business requirements. The output channels of the detection head are computed as:
\begin{equation}
  \mathrm{Filters} = (\mathrm{Classes} + 5) \cdot 3
\end{equation}
For single-class detection ($\text{Classes}=1$), the output channel is set to 18. This pruning reduces irrelevant category computation and parameter redundancy, compressing model weights to approximately 33 MB. Experiments show that this scheme provides faster loading and better memory efficiency on Raspberry Pi 5 and is therefore more suitable for long-term online operation \cite{ref6,ref8}.

\subsection{Training Strategy}
Training uses an input resolution of 416x416, an initial learning rate of 0.001, and step-wise decay at around iterations 4000 and 4500 (a 10x decrease each time), with a total of more than 5000 iterations. The training curve shows that loss decreases rapidly from about 10.0 to below 2.0 in the first 1500 iterations, enters a stable convergence region around iteration 3500, and further decreases to about 0.6 after two learning-rate decays.

\begin{figure}[H]
\centering
\includegraphics[width=0.8\textwidth]{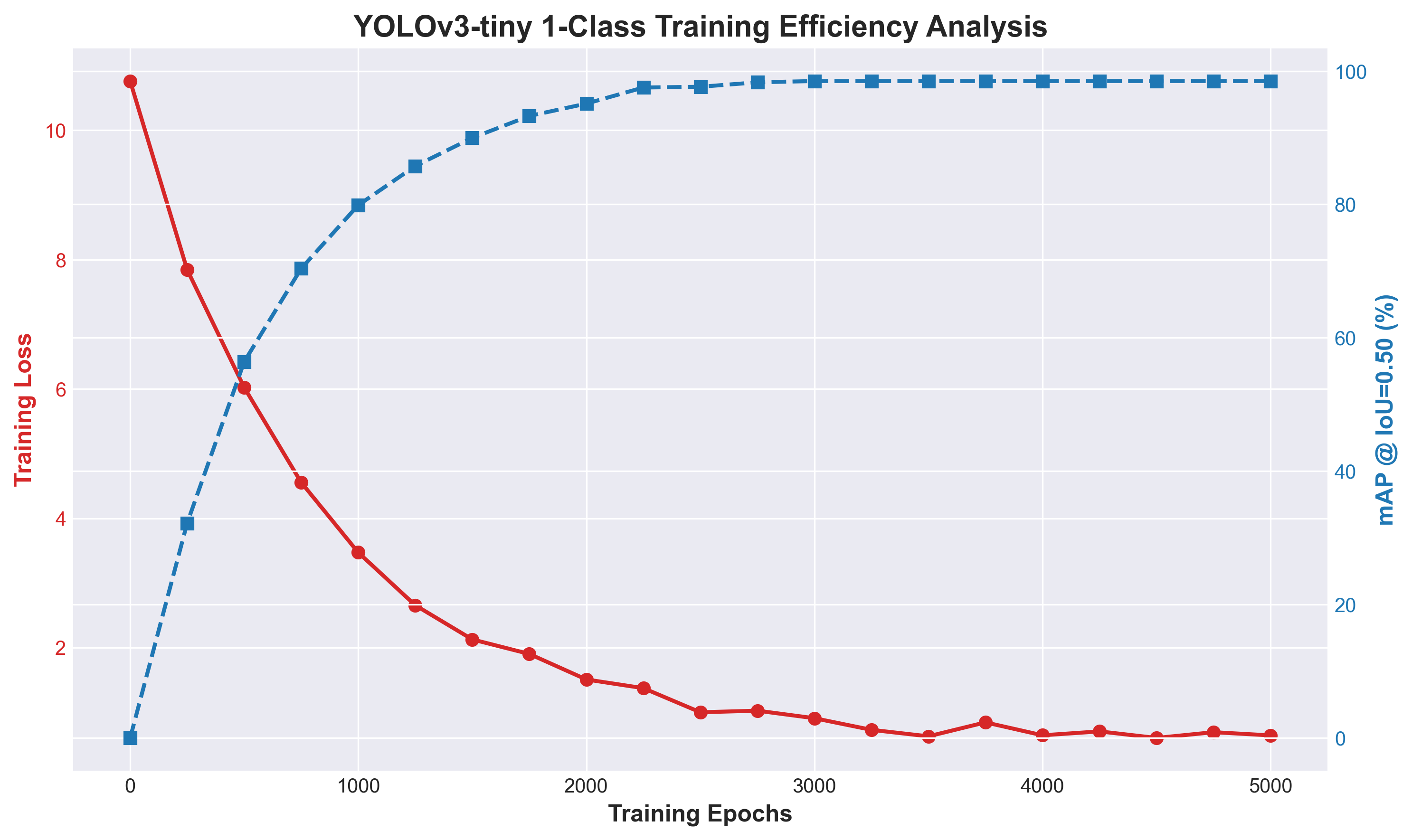}
\caption{Training-Efficiency Analysis of Single-Class YOLOv3-tiny}
\label{fig:training_efficiency}
\end{figure}

As illustrated in Fig. \ref{fig:training_efficiency}, after introducing single-class feature matching and lightweight architectural pruning, training efficiency improves significantly. Because the overall parameter scale is reduced and the task target is more focused, the loss curve drops rapidly in the initial stage. After adaptive learning-rate reductions around iterations 4000 and 4500, the model effectively exits local oscillations and smoothly enters a low and stable convergence regime. Correspondingly, the core metric mAP@0.5 steadily rises and finally reaches 96.5\%--98.2\%, indicating that the optimized model achieves both high precision and strong convergence and generalization performance in the single-class parking-detection task.

\subsection{Deployment Parameters and Runtime Strategy}
To reduce sustained inference pressure on edge-side compute and power, the system adopts a combined “event-triggered + periodic verification” strategy: detection runs every 5 s in idle-space states and every 10 s in occupied-space states; when the infrared trigger threshold is met, the visual pipeline can be awakened in advance. This mechanism prioritizes compute resources for high-risk intervals while reducing ineffective inference under guaranteed timeliness.

\section{Experimental Design and Result Analysis}
\subsection{Evaluation Dimensions}
The system is evaluated from four dimensions: recognition performance (mAP@0.5 and false-detection behavior), real-time performance (single-frame edge inference latency), system stability (heartbeat and alarm reporting stability), and energy performance (power distribution under different operating states). A two-stage protocol of “offline training evaluation + online joint debugging verification” is adopted to ensure consistency between algorithmic indicators and engineering behavior.

\subsection{Recognition Performance and Real-Time Characteristics}
On Raspberry Pi 5, for 416x416 images, single-frame inference latency after forward pass and NMS is approximately 600--850 ms. Combined with the polling strategy of every 5 s for idle spaces and every 10 s for occupied spaces, the system stably covers parking-state update requirements. With joint confidence-threshold and NMS constraints, duplicated boxes and low-quality proposals are significantly reduced, and the false-positive rate remains within an engineering-acceptable range. Recent performance evaluations of YOLO models on Raspberry Pi platforms have reported similar inference latency ranges, with optimized deployments achieving real-time operation through careful model selection and hardware-aware optimization \cite{ref20}.

\subsection{Effectiveness of Multi-Sensor Fusion}
Using only vision leads to instability in nighttime and complex-light conditions, while using only infrared tends to produce false alarms in close-range dead zones. After introducing infrared non-monotonic curve correction, historical reading smoothing and selection, and high-pressure critical-distance biasing, distance-estimation stability improves. Further integrating MPU6050 impact and tilt information significantly enhances decision consistency in scenarios such as non-vehicle disturbances, low-light false detections, and device posture anomalies \cite{ref2,ref11}. Quantitative evaluations of infrared sensors in parking systems have demonstrated that fusion with other modalities substantially improves detection reliability compared to single-sensor approaches \cite{ref14}.

\subsection{Ablation Analysis}
To quantify module contributions, comparisons are conducted across a vision baseline, a vision-plus-infrared-trigger configuration, and a full vision-infrared-inertial configuration. Results show that the infrared trigger mechanism significantly reduces ineffective inference and improves overall energy behavior, while the inertial fallback mechanism effectively suppresses missed detections in low-light and collision scenarios. In addition, infrared curve correction reduces misjudgments in close-range intervals and improves state-transition stability. These ablation results demonstrate clear modular gains of the proposed fusion framework, consistent with progressive optimization in systems engineering.

\subsection{Power Consumption Evaluation}
The system adopts hierarchical power-state transitions. In standby, baseline power is 0.8 W plus 0.12 W infrared standby, for a total of about 0.92 W. During triggered recognition, approximately 1.2 W is added by the camera and 1.8 W by YOLO inference. Compared with the conventional always-on inference mode (about 4.02 W), the optimized system maintains low power in about 80\% of idle periods, reducing average power to around 1.02 W (about 74\% energy savings) and lowering the risk of thermal throttling.

\begin{figure}[H]
\centering
\includegraphics[width=0.8\textwidth]{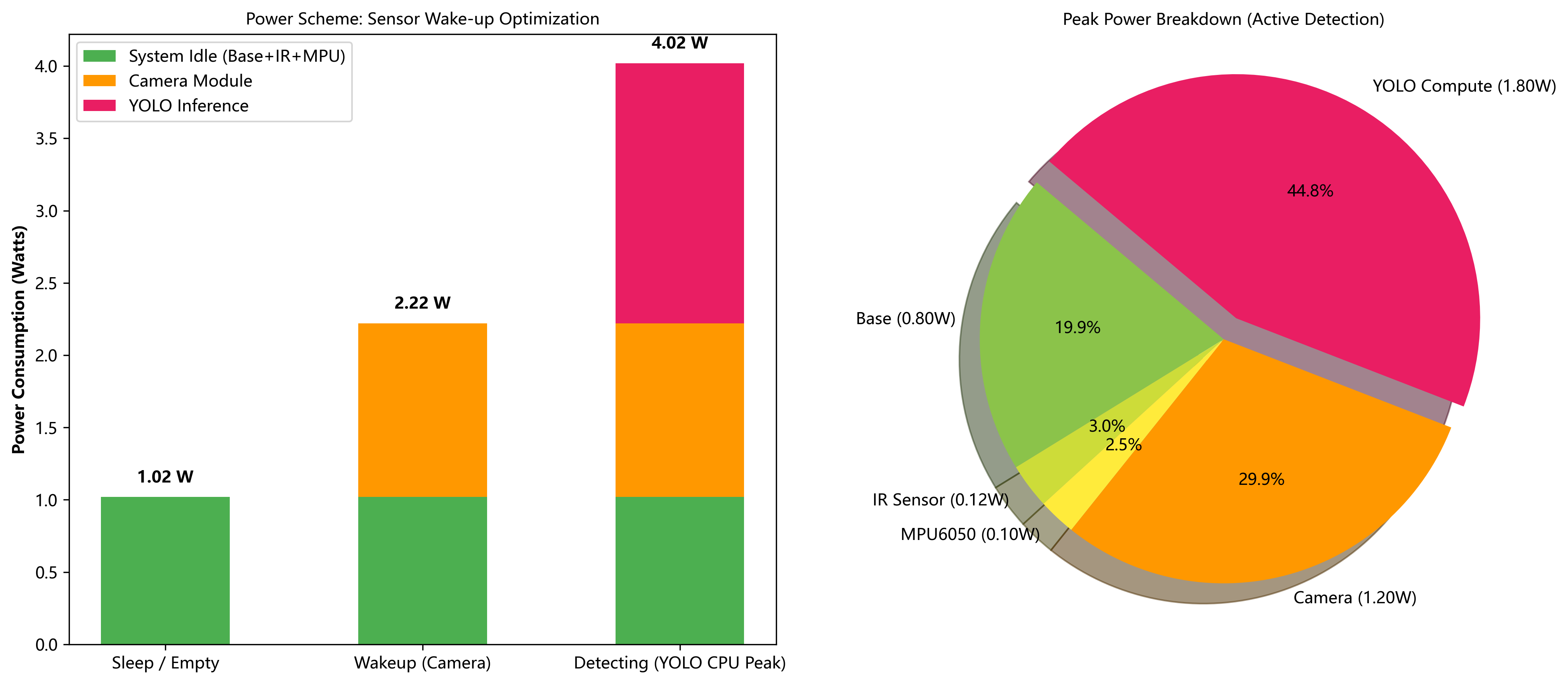}
\caption{Power Distribution Under Different Device Operating States}
\label{fig:power_consumption}
\end{figure}

As shown in Fig. \ref{fig:power_consumption}, the effectiveness of hierarchical power-state transition is evident. When spaces are idle or stably occupied, conventional always-on image streaming can easily cause overload and high energy use. In the proposed method, the system remains in low-power infrared standby for most periods, and only when significant distance changes are detected do the camera and edge YOLO module activate briefly for confirmation. This “trigger — confirm — sleep” mechanism keeps the device in an energy-saving low-load region for roughly 80\% of the time, reducing average power by about 74\% relative to the traditional setting, and thereby improving operating endurance and mitigating overheating and throttling risk in thermally constrained environments.

\subsection{Discussion of Results}
Overall, the proposed approach achieves coordinated optimization in three aspects: high-accuracy single-class recognition, practical edge-side latency, and low-power system operation. Compared with methods focusing only on algorithmic accuracy, this work emphasizes end-to-end engineering operability; compared with methods focusing only on power saving, it preserves high recognition reliability through a fusion state machine. These findings indicate that business-oriented model pruning and multi-source information fusion are practical and valuable technical paths for parking-barrier scenarios.

\section{System Integration and Business Coupling}
\subsection{Communication and Data Protocol}
Edge nodes send JSON messages to the gateway through LoRa serial links. Fields include parking-space ID, terminal ID, vehicle confidence, occupancy reason (infrared occlusion, visual confirmation, or collision fallback), sensor distance, tri-axis posture, and power information. This message design supports backend rule-engine decisions and also enables rapid troubleshooting from the operations side, ensuring state traceability and business interpretability.

\subsection{Backend-Frontend Collaborative Performance}
In multi-space joint debugging tests, the Spring Boot backend stably processes heartbeat packets and alarm events, while the frontend provides hierarchical visual display for states such as “vehicle parked in,” “illegal parking,” and “tilt damage.”

\begin{figure}[H]
\centering
\includegraphics[width=0.8\textwidth]{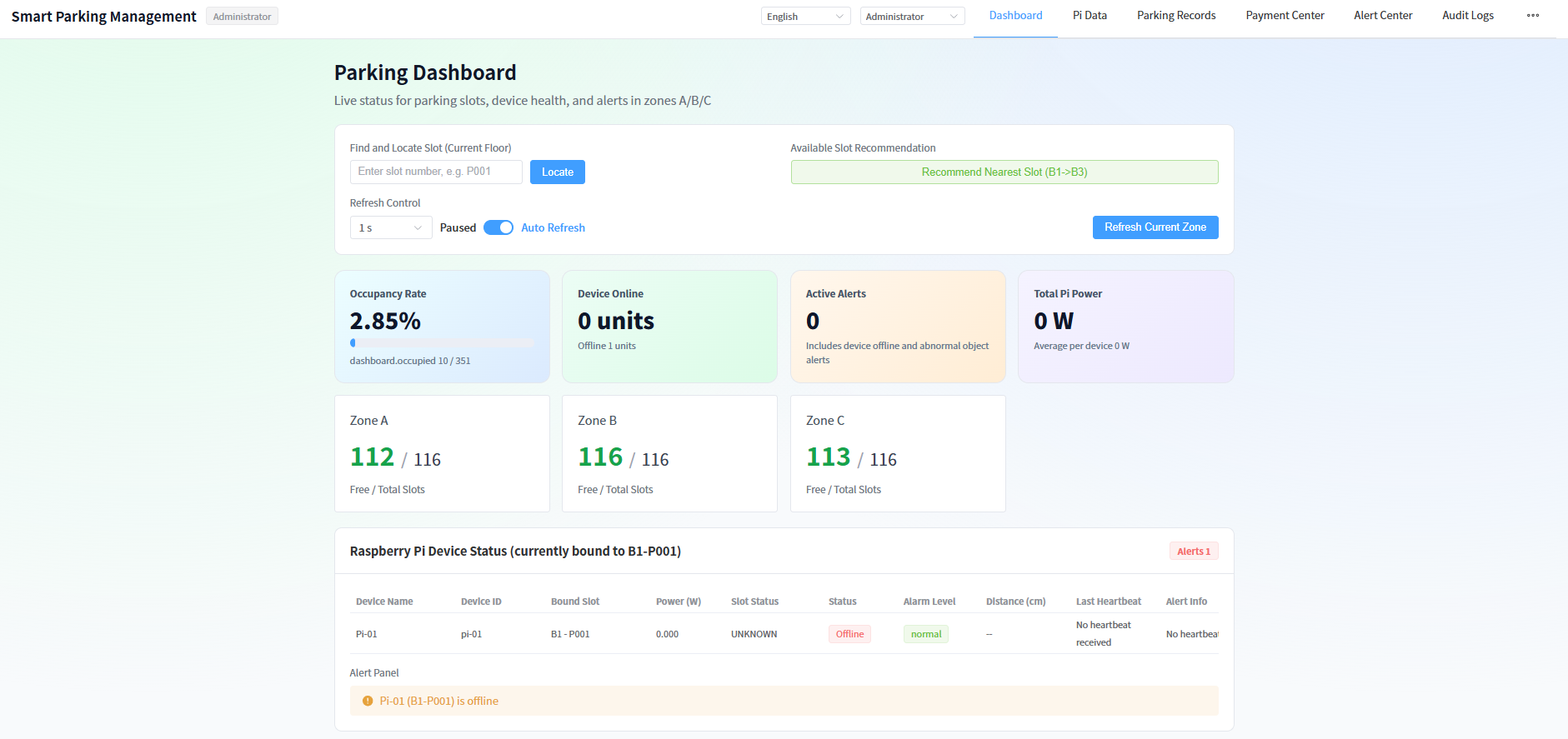}
\caption{Parking Dashboard}
\label{fig:dashboard}
\end{figure}

\begin{figure}[H]
\centering
\includegraphics[width=0.8\textwidth]{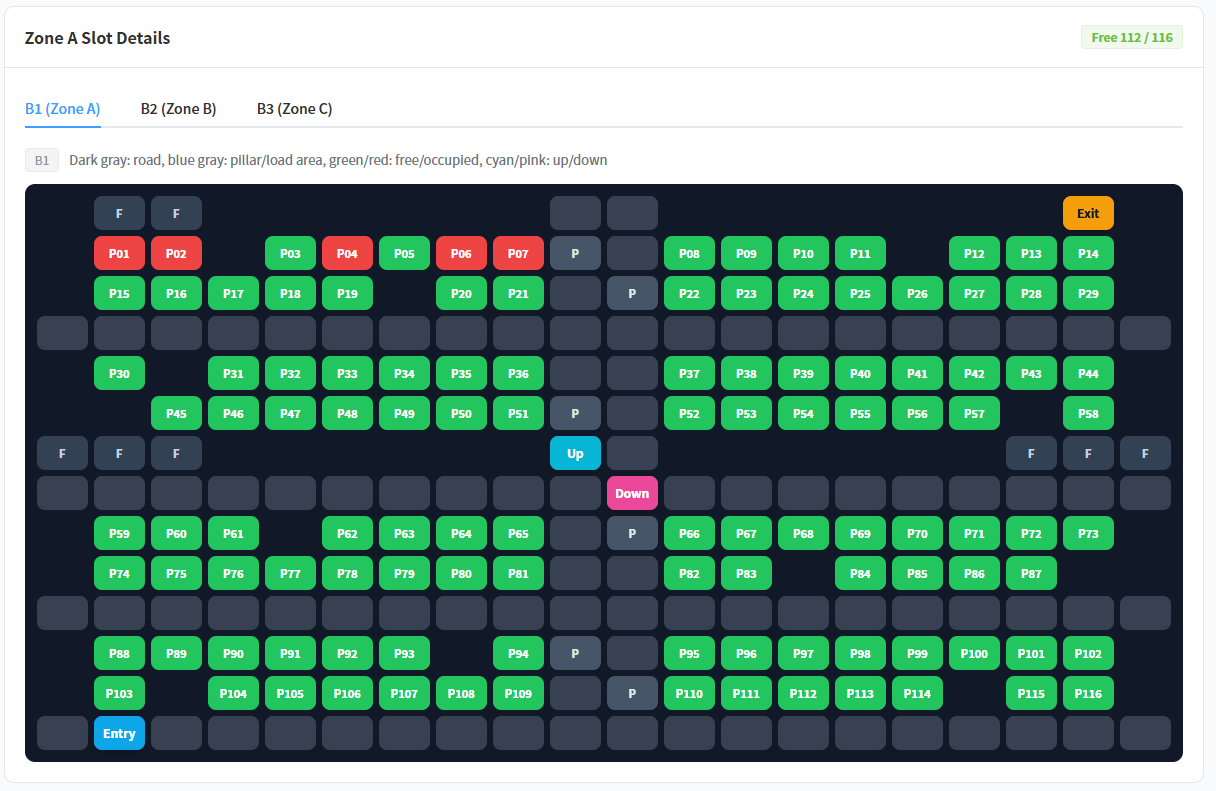}
\caption{Parking-Space Distribution Map}
\label{fig:parking_distribution}
\end{figure}

As shown by the dashboard in Fig. \ref{fig:dashboard} and the parking-space distribution visualization in Fig. \ref{fig:parking_distribution}, device-state data parsed by the cloud engine are transformed into intuitive displays in the management layer. The dashboard summarizes high-frequency daytime turnover hotspots, overall resource-utilization fluctuations, and historical fault warnings, enabling macro-level system health assessment. The parking-space distribution map uses color-threshold highlighting for micro-level terminal-state projection (e.g., red for “occupied or abnormal interception,” green for “available”). Joint debugging, pressure testing, and engineering networking results confirm that this technical framework provides fast response and stable logic across the full pipeline of data acquisition, IoT cloud transmission, management-layer synchronization, and alarm display, satisfying stringent functional requirements for large-scale commercial deployment of modern parking barriers \cite{ref4,ref7}.

\subsection{Engineering Maintainability Analysis}
From an operations and maintenance perspective, the proposed system has several advantages. The edge nodes are clearly modularized, facilitating incremental deployment by parking-space scale. The cloud side supports node health monitoring and offline alarms via heartbeat mechanisms. The front end provides state stratification and reason explanation, reducing manual inspection cost. These characteristics improve long-term availability in real deployment scenarios.

\section{Conclusion and Future Work}
To meet engineering deployment requirements of intelligent parking barriers, this paper proposes and implements a system combining lightweight visual detection, multi-sensor fusion, and edge-cloud collaborative management. The results indicate that single-class YOLOv3-tiny pruning achieves high accuracy with acceptable latency on Raspberry Pi 5, demonstrating edge deployment feasibility. The infrared-vision-inertial asymmetric fusion mechanism significantly improves robustness of state determination in complex scenarios. The hierarchical power-state transition strategy effectively reduces average power and alleviates thermal load. Edge-cloud joint debugging validates closed-loop consistency from perception to business processing and then to visualization. Overall, the proposed solution achieves a favorable balance among recognition performance, system stability, energy control, and maintainability.

Future work may proceed in several directions. Larger-scale datasets across seasons and extreme weather conditions can be built to improve model generalization \cite{ref9,ref12}. Adaptive thresholds and online calibration mechanisms can be introduced to support parking-space-level personalized configuration. New-energy power supply and edge model distillation can be combined to further improve endurance and real-time performance \cite{ref10}. Long-term tests in large parking lots can be conducted to quantify improvements in operation and maintenance efficiency and false-alarm cost. Additionally, emerging frameworks that integrate YOLO with quantized large language models on edge devices offer promising directions for enhancing system intelligence and natural interaction capabilities \cite{ref20}. The integration of parking edge computing with vehicular edge computing paradigms also presents opportunities for broader urban mobility optimization.

\section*{Acknowledgment}
Yuwen Zhu wrote the manuscript. Yuwen Zhu and Feiyang Qi implemented the system and conducted the experiments. Zhengzhe Xiang provided the hardware and contributed to the conceptualization. This work was supported in part by the College Student Innovation and Entrepreneurship Training Program of Hangzhou City University.

\bibliographystyle{IEEEtran}
\bibliography{ref}

\end{document}